\documentstyle[onecolumn]{mn}
\newcommand{\mincir}{\raise 
-2.truept\hbox{\rlap{\hbox{$\sim$}}\raise5.truept
\hbox{$<$}\ }}
\newcommand{\magcir}{\raise 
-2.truept\hbox{\rlap{\hbox{$\sim$}}\raise5.truept
\hbox{$>$}\ }}
\newcommand{\minmag}{\raise-2.truept\hbox{\rlap{\hbox{$<$}}\raise 
6.truept\hbox
{$>$}\ }}
\newcommand{\be}{\begin{equation}}
\newcommand{\ee}{\end{equation}}
\newcommand{\ba}{\begin{eqnarray}}
\newcommand{\ea}{\end{eqnarray}}
\newcommand{\brr}{\begin{array}}
\newcommand{\err}{\end{array}}
\newcommand{\bc}{\begin{center}}
\newcommand{\ec}{\end{center}}
\newcommand{\br}{\mbox{\bf r}}
\newcommand{\bv}{\mbox{\bf v}}

\newcommand{\bg}{\mbox{\bf g}}

\newcommand{\bk}{\mbox{\bf k}}

\newcommand{\hm}{\,h^{-1}{\rm Mpc}}
\newcommand{\vel}{\,{\rm km\,s^{-1}}}

\def\ifm#1{\relax\ifmmode#1\else$\mathsurround=0pt #1$\fi}
\def\kms{\ifmmode\,{\rm km}\,{\rm s}^{-1}\else km$\,$s$^{-1}$\fi}
\def\hmpc{\ifmmode\,{\it h }^{-1}\,{\rm Mpc }\else $h^{-1}\,$Mpc\,\fi}

\input epsf
\def\\{\hfill\break}

\def\etal{{\it et al.\ }}

\def\cf{{\it cf.}}

\def\ifm#1{\relax\ifmmode#1\else$\mathsurround=0pt #1$\fi}
\def\kms{\ifmmode\,{\rm km}\,{\rm s}^{-1}\else km$\,$s$^{-1}$\fi}

\def\ltsima{$\; \buildrel < \over \sim \;$}
\def\lsim{\lower.5ex\hbox{\ltsima}}
\def\gtsima{$\; \buildrel > \over \sim \;$}
\def\gsim{\lower.5ex\hbox{\gtsima}}
 
\def\pmb#1{\setbox0=\hbox{#1}%
 \kern-.025em\copy0\kern-\wd0
 \kern.05em\copy0\kern-\wd0
 \kern-.025em\raise.0433em\box0}
\def\vv{\pmb{$v$}}

\def\vr{\pmb{$r$}}

\def\v0{\pmb{$0$}}

\title [Local Group Acceleration]{Likelihood Analysis of the Local Group Acceleration}
\author[I. Schmoldt \etal]
{ I. Schmoldt$^{1}$, E. Branchini$^{2}$, L. Teodoro$^{2}$,\\
\vspace{-1mm}\\ 
{\LARGE G. Efstathiou$^{3}$, C.S. Frenk$^{2}$, O. Keeble$^{5}$, 
S. Maddox$^{3}$, S. Oliver$^{5}$, }\\ 
\vspace{-1mm}\\ 
{\LARGE M. Rowan-Robinson$^{5}$, W. Saunders$^{4}$, W. Sutherland$^{1}$, H. Tadros$^{6}$, 
S.D.M. White$^{7}$ }\\ 
$^1$ Department of Physics, University of Oxford, Keble Road, Oxford
OX1 3RH, UK \\
$^2$Department of Physics, University of Durham, South Road, Durham DH1
 3LE, UK \\
$^3$ Institute of Astronomy, University of Cambridge, Madingley Road,
Cambridge CB3 OHA, UK\\
$^4$ Institute for Astronomy, University of Edinburgh, Blackford Hill,
Edinburgh EH9 3JS, UK  \\
$^5$ Imperial College of Science, Technology, and Medicine, Blackell
Laboratory, Prince Consort Road, London SW1 2EZ, UK\\
$^6$ Department of Physics, University of Sussex, Falmer, Brighton BN1
9QH, UK \\
$^7$ Max Planck Institut f\"{u}r Astrophysik,
Karl-Schwarzschild-Stra{\ss}e 1, 85740 Garching, Germany}

\begin{document}
\maketitle

\begin{abstract}
We compute the acceleration on the Local Group using 
11206 IRAS galaxies from the recently completed all-sky PSC$z$ redshift 
survey.
Measuring the acceleration vector in redshift space
generates systematic uncertainties due to the redshift space
distortions in the density field. We therefore assign galaxies to their
real space positions by adopting a non-parametric model for the
velocity field that solely relies on the linear gravitational instability 
and linear biasing hypotheses. Remaining systematic contributions to
the measured acceleration vector
are corrected for by using
PSC$z$ mock catalogues from N-body experiments.

The resulting acceleration vector points $\sim 15^{\circ}$ away
from the CMB dipole apex, with a remarkable alignment between small
and large scale contributions. A considerable fraction
($\sim 65 \%$) of the measured  acceleration
is generated within 40 \hmpc with a non-negligible
contribution from scales between 90 and 140 \hmpc
after which the acceleration amplitude seems to have converged.
The local group acceleration from PSC$z$ appears to be consistent with  
the one determined from the IRAS 1.2 Jy galaxy catalogue once the 
different contributions from shot noise have been taken into account.
The results are consistent with the gravitational
instability hypothesis and do not indicate any strong deviations from
the linear biasing relation on large scales.

A maximum-likelihood analysis of the cumulative PSC$z$ dipole
is performed within a radius of 150 \hmpc in which we account for
nonlinear effects, shot noise and finite sample size.
The aim is to constrain the $\beta=\Omega^{0.6}/b$ parameter and the 
power spectrum of density fluctuations.
We obtain $\beta= 0.70^{+0.35}_{-0.2}$ 
at 1 $\sigma$ confidence level.

The likelihood analysis is not very
sensitive to the shape of the power spectrum due to the rise in
the amplitude of the dipole beyond 40 \hmpc and due to the increase in
shot noise on large scales. There is however a weak
indication that within the framework of CDM models the observed Local
Group acceleration implies some excess power on large scales.

\end{abstract}


\newpage
\section{Introduction}
%1
\label{sec:intro}

If the Cosmic Microwave Background [CMB] defines a cosmological frame,
then the dipole pattern observed in its temperature is a direct measure,
via Doppler shift, of the Local
Group [LG] velocity: $v_c=627 \pm 22$ \kms towards
$(l,b)=(276^{\circ}\pm3^{\circ},30^{\circ}\pm 2^{\circ})$ 
as inferred from the 4-year-COBE data (Lineweaver \etal 1996). 
The best reason to prefer a velocity interpretation 
to more exotic scenarios (e.g. the Tolman-Bondi cosmological model of
Paczynski \& Piran 1990, where the dipole moment of the CMB is of 
cosmological origin) 
is the remarkable alignment of the dipole with the gravitational
acceleration vector measured at the LG location. This gravitational
acceleration is inferred from the distribution of luminous objects in
our cosmological neighbourhood 
and is expected to align with the dipole vector within the
Gravitational Instability (Peebles 1980) and linear biasing
framework in which the observed structures in the universe
grew via gravitational instability from initial small fluctuations
in the mass density field. 
In the linear regime, i.e. on scales where the mass density contrast,
$\delta_{\rho}$, is small, we also expect a direct proportionality
between the amplitudes of the two vectors.
The proportionality constant is a measure of the so-called $\beta(=
\Omega_m^{0.6}/b)$ parameter, where $\Omega_m$ represents the density and
$b$, the bias parameter, relates the mass 
density contrast, $\delta_{\rho}$, to the 
fluctuations in the number density of the luminous objects, $\delta_g
=b \delta_{\rho}$.
A comparison of the observed LG velocity with the gravitational 
acceleration is therefore a direct 
measure of $\beta$.

The many different estimates of the LG acceleration (see
Strauss and Willick (1995) for a complete review) are partly
contradictory.
Most of the measurements agree in showing a small ($\le 30^{\circ}$)
misalignment between the CMB and the LG dipoles.
The amplitudes, however, are highly sensitive to the objects
used as mass tracers.
As reported by Strauss (1997), the LG acceleration as calculated
using galaxies extracted from the  IRAS 1.2 Jy redshift
survey (Fisher \etal 1995) and from the Optical Redshift Survey [ORS]
(Santiago \etal 1995, 1996) seems to receive little contribution
from scales larger than $40 \hmpc$.
Other analyses carried out using a
deeper but sparser redshift survey of IRAS galaxies, known as the QDOT 
catalogue,
have shown a non-negligible contribution from scales up to $\sim 100
\hmpc$ (Rowan--Robinson \etal 1990).
Clusters of galaxies, both optically and X-ray selected, can be used to
probe very large scales because of their large luminosity. The LG
gravitational acceleration generated by their spatial distribution
also provides evidence of non-negligible contributions from depths 
up to $\sim 150 \hmpc$ (Scaramella, Vettolani \& Zamorani 1991, Plionis
\& Valdarnini 1991,  Plionis \& Kolokotronis 1998).

Such an apparent dichotomy may arise from the limited depth of the ORS and
IRAS 1.2 Jy catalogues, from the sparse sampling in the QDOT and
cluster catalogues, or from inhomogeneity and observational biases that 
may affect the different samples.

In this paper, we use
the recently completed PSC$z$ survey of IRAS galaxies. 
This is a homogeneously selected all-sky catalogue which combines the
depth of the QDOT with the dense sampling of the 1.2 Jy catalogue and
therefore represents the ideal tool for investigating the large scale
contribution to the LG gravitational acceleration. Our main concern
is the combination of information on the growth of the
dipole with other measurements of the local velocity fields (mainly
from POTENT data) in a likelihood analysis and we therefore do not
use the catalogue to its full extent. Rowan--Robinson \etal (1998)
have done a careful analysis of the growth of the dipole on the
largest scales using the same catalogue out to much larger distances.

Even with an ideal
flux-limited catalogue, however, other effects 
(shot noise, finite depth of the sample,  use of the linear 
approximation, incomplete sky coverage,
and redshift space distortions) cause the LG acceleration to 
differ from the observed LG velocity, making it impossible to determine $\beta$
by direct comparison.
In this work, we adopt an alternative two--step procedure in which:
\begin{itemize}
\item the LG cumulative acceleration is computed from the data 
and its random and systematic errors are evaluated using PSC$z$ mock
catalogues extracted from  N-body simulations. 
These catalogues are also used to statistically correct the measured
PSC$z$ dipole amplitude for systematic errors and to obtain a
preliminary estimate of $\beta$ from the data.
\item the resulting bias-free gravitational acceleration is then
analysed using a likelihood approach similar to the one developed 
by Strauss \etal (1992) [S92 hereafter]. This will yield an estimate of 
$\beta$ within the framework of a given cosmology and 
a way of evaluating errors.
\end{itemize}

The outline of this paper is as follows:
In \S2 we give a short theoretical introduction.
\S3 is devoted to computing the LG acceleration from the data and
to estimating numerical biases and errors with the help of mock catalogues.
\S4 contains a likelihood analysis (theory and application) of the
results, and conclusions follow in \S5.

\section{Theoretical Background}
\label{sec:theory}

In linear theory, the peculiar velocity, $\bv$, 
of a given galaxy is directly proportional to
the gravitational acceleration, $\bg$, 
caused by the surrounding matter. 
The  two vectors are therefore parallel and related by
\begin{equation}
\bv(\br) = \frac{H_{\circ}f}{4 \pi G} \bg(\br),
\label{vd}
\end{equation}
where $f \sim \Omega_m^{0.6}$.
Hence, if the peculiar velocity is known, $\Omega_m$ can be
determined by calculating the gravitational acceleration from the
observed mass distribution, where $\Omega_m$ is the density of all
matter in units of the critical density.

During the past few years, particular attention has been paid to
applying eqn. (\ref{vd}) to 
compute the acceleration acting on 
the Local Group of galaxies. The Local
Group velocity in the CMB frame
is accurately known from the CMB dipole
anisotropy
\footnote{Note that both the CMB dipole and the gravitational
acceleration caused by the surrounding matter are often 
referred to as `dipoles'. In this paper we will obey this 
convention except when distinguishing between
the  `CMB velocity' ($\bv_c$), which is inferred from the CMB dipole
anisotropy, and the `measured velocity' ($\bv_r$) inferred from 
the gravitational acceleration.}
while the gravitational acceleration vector can be measured
from redshift surveys which give an estimate of the three-dimensional
galaxy density field. The vector $\bg$ is related to the mass density
field by
\begin{equation}
\bg(\br) = G  \int \delta_{\rho}(\br\prime)
\frac{\br\prime }{\mid \br\prime \mid^{3}}
d^3\br\prime
\label{gg}
\end{equation}
where $\delta_{\rho}(\br)$ is the density contrast at comoving coordinate
$\br$. 
Since we sample the density contrast of the galaxy rather than
of the matter distribution, we need to postulate some relation between
the two. In this work, we assume linear biasing, in which 
$\delta_g=b\delta_{\rho}$, such that we can shift from galaxy to matter
density simply by dividing by the bias factor $b$. Eqn. (\ref{vd})
then becomes
\begin{equation}
\bv= \frac{H_{\circ} \beta}{4 \pi} \int d^{3} \br \prime
\frac{\br\prime}{\mid \br\prime\mid^{3}}
\delta_g(\br\prime), 
\label{lgvelb}
\end{equation}
where $\beta=f/b$. 

In principle, $\beta$ can easily be obtained by comparing $|\bv_r|$
and $|\bv_c|$. In practice, however, the
right hand side of eqn. (\ref{lgvelb}) is difficult to evaluate from
redshift surveys because of several effects:

\begin{itemize}
\item Finite Sample Size: The integral of
equation \ref{lgvelb} 
extends over all space. When using a
redshift survey, however, this necessarily turns into a sum over a finite
number of galaxies. The volume sampled extends to the maximum depth of 
the survey and excludes possible unsurveyed regions.
Evaluating eqn. (\ref{lgvelb}) in a finite survey volume results in
a discrepancy between $\bv_r$ and $\bv_c$.
\item Shot Noise: At large radii, the sampling of the
galaxy distribution becomes more and more sparse, which leads to an
increase in the so-called shot noise error. In this work, the 
shot noise contribution is evaluated following the S92 approach, 
which also accounts for the uncertainty in the galaxy mass function,
i.e. the uncertainty associated with the assumption that all galaxies
have the same mass.
\item Redshift Space Distortions: The surveys do not give full
three-dimensional positions for the galaxies but rather angular
positions and redshifts. Redshifts are related to distances via
Hubble's law but only in the limit of very large distances. For
nearby objects the peculiar velocities of the galaxies will add a
significant contribution to the redshift that systematically distorts any
density field calculated  in redshift space (Kaiser 1987, Kaiser and
Lahav 1988). We attempt to
minimise redshift space distortions
by reconstructing galaxy
positions in real space from their redshifts.
\item Nonlinear Effects:
Eqn. (\ref{lgvelb}) is not valid over the entire survey
volume since linear theory breaks down in high density
regions. 
Nonlinear contributions to peculiar velocities 
add incoherently to the vector, spoiling 
the linear relationship between velocity and acceleration vectors.
\end{itemize}

All these effects have to be accounted for before a meaningful
statement can be made about the value of $\beta$ calculated from a
comparison of $\bv_r$ and $\bv_c$.
In what follows, we will use a likelihood
analysis that models the non-linear effects 
and the finite survey volume. The shot noise
contribution will be calculated directly from the data 
while a self-consistent dynamical algorithm is adopted
to model the errors introduced when minimising redshift space
distortions.

\section{The Local Group Acceleration}
\label{sec:Dipole}
In this section, we present the dataset used, our methods for
calculating the dipole and for evaluating the associated errors and
finally the results of these procedures. By using simulated redshift
catalogues, we hope to be able to correct the calculated dipole for
all those errors that cannot be accounted for in the likelihood analysis.

\subsection{The Dataset}
\label{sec:dataset}

Our dataset comes from the recently completed
IRAS PSC$z$ catalogue 
-- a redshift survey of some 15,500 galaxies detected in 
the IRAS Point Source Catalogue.  
The PSC$z$ catalogue contains almost every galaxy in the IRAS
PSC with 60 $\mu m $ flux ($f_{60}$) larger than 0.6 Jy
for which the redshift has been measured. 
Only sources from PSC with $f_{60}>0.5 f_{25}$ (to exclude stars) and
with $f_{100}<4 f_{60}$ (to exclude most of the galactic cirrus)
have been included in the catalogue.
Our subsample contains 11206 PSC$z$ objects within
200 \hmpc, with PSC fluxes at 60 $\mu m >0.6$ Jansky, and
with positive galaxy identifications. 
Regions not surveyed by IRAS (two thin
strips in ecliptic longitude and the area near the galactic 
plane defined by a V-band extinction of $> 1.5$ mag.) 
are excluded from the catalogue, which therefore 
covers $\sim 84 \%$ of the sky. 

We also consider 
a catalogue of galaxies extracted 
from the 1.2 Jy redshift survey (Fisher \etal 1995)
which contains all the objects with $f_{60}>1.2$ Jy,
also within 20,000 \kms.
The resulting  1.2 Jy subsample has $\sim 88 \%$
sky coverage  and contains 4626 galaxies.

\subsection{Minimising Redshift Space Distortions} 
\label{sec:reconstruction}

As outlined by Kaiser (1987) and Kaiser and Lahav (1988), redshift space 
distortions can
modify the amplitude of the galaxy dipole, especially  when it is
measured from a flux-limited sample. To minimise this effect,
we compute the galaxies' peculiar velocities and real space positions 
iteratively
starting from the galaxy distribution as seen in the redshift space. 
A detailed methodological description
can be found in Branchini \etal (1998), here we just summarise the main 
concepts. The method described is very similar to that developed 
by Yahil \etal (1991).

We assume linear GI and LB and look for an iterative solution to
the system of equations
\begin{equation}
 \bv(\br) = \frac{H_0 \beta}{4 \pi} \int d^{3} \br\prime
\frac{\br\prime - \br} {\mid \br\prime - \br \mid^{3}}
\delta(\br\prime) 
\label{vr}
\end{equation}
and
\begin{equation}
\br_i = cz_i - \hat{\br} \cdot (\bv_i - \bv_c),
\label{czr}
\end{equation}
where $\bv_i$ is the peculiar velocity of the generic object $i$,
$\bv_c$ 
is the LG velocity (both velocities are measured in the CMB frame), $\hat{\vr}$
is the unit vector 
along the radial direction. Distances and velocities are 
expressed in $\kms$. The redshift $z_i$ is measured in the LG
frame.
First, we make a  guess for $\br$ (which in the 0-th iteration
coincides with the galaxy redshift),
then compute $\bv(\br)$ from eqn. (\ref{vr}),
correct to a new $\br$ by eqn. (\ref{czr}) and continue the process
until the calculated velocities $\bv_i$ converge.
This is typically achieved within 8 iterations.

There are several additions to this, designed to minimise the
uncertainties inherent in the procedure.
When computing the gravitational acceleration from 
eqn. (\ref{vr}) we smooth the density field using a Top Hat window
and a smoothing length equal to the average separation of objects
at that distance. The minimum smoothing length is 5 \hmpc to ensure
sufficient smoothing even at distances where the
interparticle separation is small.
This procedure mitigates nonlinear effects 
especially in high density environments but also 
impairs our ability to predict the 
small-scale features in velocity and density fields. 

The free  parameter $\beta$  has to be determined {\it a posteriori}
by comparing observations with model predictions.
We adiabatically increase
$\beta$ from 0.1 to 1.0 in 10 iterations,
thereby slowly increasing the effect of gravity to improve 
convergence of the velocity field (S92).
The resulting peculiar velocities scale to first order in $\beta$ which,
as we will discuss in \S 3.4, causes the galaxy dipole to be
almost independent of the $\beta$ value used in the reconstruction.

The iterative procedure uses all the galaxies within 200 \hmpc
while the mass distribution beyond is considered 
to be homogeneous, i.e. we neglect its
gravitational effect.
This assumption is justified as long as the reconstruction is performed
in the LG frame since relative peculiar velocities are not 
affected by external dipoles. Higher moments of
external force fields only affect the reconstruction reliability
in the outer regions. To avoid these edge effects we calculate peculiar
velocities and real space positions only for galaxies out to 170 \hmpc
even though material out to 200 \hmpc is considered to be
gravitating. The error analysis in \S 3.4 indicates that the combined
effect of
shot noise and reconstruction uncertainties at that distance have
already increased to over 30 \% of the measured acceleration, so to
maximise the reliability of the data, we perform the likelihood
analysis only on data measured out to 150 \hmpc.

Spurious gravitational acceleration may also arise if the 
galaxy sample does not have full sky coverage. To alleviate 
this effect, we fill the unobserved regions with synthetic 
objects using two distinct procedures:
At high ($|b|\ge 8^{\circ}$)  galactic latitude we fill the
masked areas with a uniform distribution of simulated objects having the
same average number density and selection function as the PSC$z$
galaxies. 
At lower galactic latitudes ($|b| < 8^{\circ}$, the so called `zone of 
avoidance') the masked area
is filled  with galaxies which are cloned 
from the PSC$z$ galaxy distribution in two strips above 
and below the zone of avoidance (Yahil \etal 1991), also retaining
PSC$z$ galaxies with $|b| < 8^{\circ}$ at their observed position.

Since we are using a flux-limited catalogue, the number 
density of galaxies 
within the survey volume decreases with the distance.
In the reconstruction we account for this effect
by assigning a weight to each galaxy that is proportional to the inverse 
of the selection function
at the position of the galaxy. Since the positions of the galaxies vary
during the iterations, 
we recompute the selection function at 
each step from the observed galaxy fluxes using 
a parametric maximum likelihood technique,
and the galaxy weights are updated accordingly.
For the selection function we assume the analytic form proposed 
Yahil \etal (1991).

Most of the results presented here refer to this iterative reconstruction
procedure. However, two other methods, fully described
and tested  in Branchini \etal (1998), have been implemented to check 
the robustness of the results.
The first is a grid-based version of the above iterative method 
while the second is based on the Nusser and Davis approach (1994)  
that uses the first order Zel'dovich approximation to
correct for redshift space distortions without any need for
iterations.
In what follows we will refer to these three methods as M1, M2 and
M3. As we will show in \S 3.5, the PSC$z$  dipoles reconstructed 
with the three methods are fully consistent with each other.

\subsection{PSC$z$ Mock Catalogues} 
\label{sec:mock1}

Systematic uncertainties affecting the measured PSC$z$ dipole 
have been evaluated using a series of PSC$z$ mock catalogues.
These were extracted from  
N-body simulations of CDM universes 
performed by Cole \etal (1998)
using a 345.6 \hmpc computational box.
For the present analysis we have considered 
two
of their 
cosmological models.
They are: 
\begin{itemize}
\item an $\Omega_m=1$ CDM model with $\Gamma=0.25$ 
\item an $\Omega_{\Lambda}=0.7$, $\Omega_m=0.3$ CDM model
\end{itemize}
These two are flat CDM cosmologies normalised to reproduce the
local abundance of clusters as explained in Eke, Cole, and Frenk (1996):

\noindent

$\sigma_8=0.52\Omega_m^{-0.46+0.10\Omega_m} \ {\rm if} \ 
\Omega_{\Lambda}=0$, and

\noindent

$\sigma_8=0.52\Omega_m^{-0.52+0.13\Omega_m} \ {\rm if} \ 
\Omega_{\Lambda}=1-\Omega_m$.

\noindent

The velocity fields of the N--body models have been smoothed with a top
hat function of 1.5 \hmpc to obtain
pairwise peculiar velocities of $\sim 250$ \kms at 1 \hmpc, 
a value close to that observed
in the galaxy distribution (Guzzo \etal 1997, Strauss, Ostriker 
and Cen  1998).
For each of the two models we produce 10 mock catalogues:
We impose $b=1$, a value compatible
with the one commonly accepted for IRAS galaxies, in order to identify
galaxies with particles in the
simulations. The velocity field in each catalogue mirrors the
conditions in the local universe: the central LG-like observer 
has a 
peculiar velocity of
${\bf v}_{c}= 625 \pm 25$ \kms, the LG
velocity residuals $\langle (v_i-v_c)^2 \rangle^{0.5}$
within $5 \hm$ (sometimes called ``shear'') are 
smaller than 200 \kms and the fractional overdensity within the same
region is in the range $-0.2 < \delta < 1.0 $ (Brown and Peebles 1987).
Each catalogue has a depth of $170 \hm$ (since our velocity analysis
of the PSC$z$ will only extend out to that distance)
and the coordinate system is rotated such that the 
observer's velocity vector points towards the observed CMB apex.
Since the number density in the simulations
($\sim 0.039$ $h^3$  Mpc$^{-3}$)
is smaller than 
the one given by the PSC$z$ selection function 
within 10.9 \hmpc, 
we are forced to semi-volume limit our catalogues at that radius.
At greater distances we use a Monte Carlo rejection 
technique to force the
simulated galaxy population to obey the
$N(z)$ distribution of PSC$z$ galaxies.
A random flux consistent with the selection 
function is then attributed to each galaxy. 
Finally, we mask the same areas as in the PSC$z$ catalogue to mimic the
incomplete sky coverage.
A similar set of catalogues has been produced to mimic the 
galaxy distribution in the IRAS 1.2 Jy
subcatalogue (\S 3.1).
Given the smaller number density of 1.2 Jy galaxies (Fisher \etal
1995), the 1.2 Jy
mock catalogues are semi-volume limited to a smaller radius of $7.8 \hm$.

\begin{figure}
\parbox{15 cm}{%
\epsfxsize=\hsize\epsffile{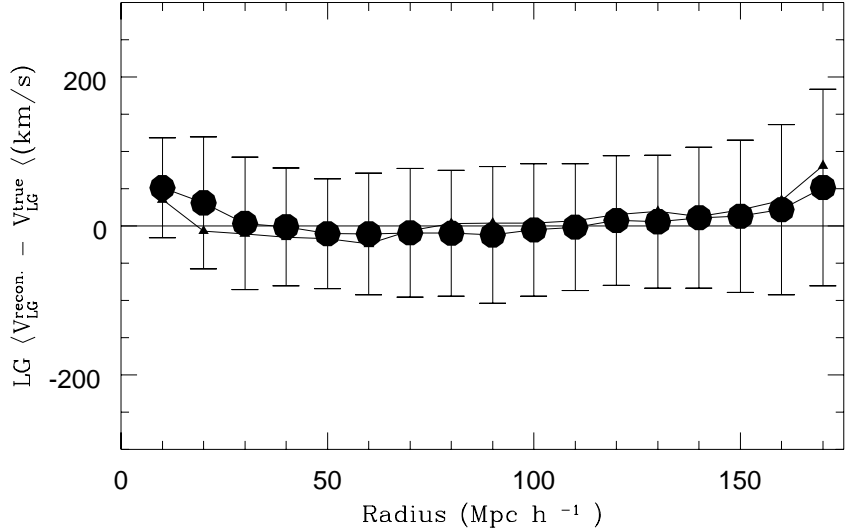}}
\caption{Error estimates from the mock catalogues.
Black dots show the discrepancy between the true and the ``observed'' 
PSCz dipole, averaged over different mock catalogues at different radii.
Error bars represent the scatter around the mean
and quantify the amplitude of the random errors. 
Small triangles show similar results for the 1.2 Jy mock dipoles. 
}
\end{figure}

\subsection{Error Estimates Using Mock Catalogues} 
\label{sec:mock2}

Several different effects prevent the reconstruction procedure 
from recovering the exact real space galaxy distribution and 
a less than perfect reconstruction generates both systematic and 
random errors in the dipole. 
S92 showed that the most serious systematic bias
originates from the so called `rocket effect' (Kaiser 1987, Kaiser and 
Lahav 1988) which represents the spurious acceleration measured 
from a magnitude-limited sample of galaxies
by an observer with some peculiar velocity unconnected to the
gravitational acceleration (e.g. an
observer in a rocket).
In our case the errors in the computed LG velocity mimic
the non-gravitational acceleration.
The characteristic signature of this effect 
is a monotonically increasing spurious contribution
to the cumulative LG acceleration. 
S92 correct for this effect by modifying
their reconstruction procedure. 
Here we adopt a different strategy which uses mock catalogues 
to correct not only 
for the rocket effect but also for further possible systematic 
effects 
that arise from filling the unsurveyed areas with synthetic objects.
We consider a generic PSC$z$ mock catalogue 
and measure the gravitational acceleration at the observer's central position
from the true, all--sky galaxy distribution in real space.
We then measure the gravity vector
from the same catalogue, 
after applying the filling procedure and after iteratively moving 
galaxies to their reconstructed real space positions.
The two acceleration vectors
are similarly affected by shot noise, finite sample
size and nonlinear effects and thus any discrepancy can only
be ascribed to the intrinsic uncertainties in the reconstruction, the
filling techniques and to the rocket effect.   
For each of the mock catalogues we compute the vector difference between 
the two cumulative accelerations at different radii, thereby obtaining
an estimate for the systematic dipole error. We then average
the difference over the various catalogues of both cosmologies
to account for our ignorance of the underlying, true cosmological
model.
The mean discrepancy represents the cumulative effect 
of the various systematic errors while the dispersion around the mean
quantifies the random errors.
A similar exercise has been performed using mock IRAS 1.2 Jy 
catalogues.
Results of this procedure are displayed in figure 1. 
The points represent
the systematic error on the cumulative dipole measured 
at a given radius whereas the errorbars display
the dispersion in that error. 
Since we are showing a cumulative quantity, 
the errors on each shell are correlated, both in amplitude and 
direction.
Note
that $v_{true}$ is the dipole velocity calculated using real space
positions of galaxies, whereas $v_{rec}$ is computed using the
redshift space positions as described in \S 3.2. 
Both quantities are normalised to $\beta=1$.
Filled dots refer to PSC$z$ mocks while 
small triangles refer an analogous analysis performed 
using the simulated IRAS 1.2 Jy catalogues.

It is encouraging to note that there is virtually 
no systematic error
up to a distance of 150 \hmpc. The marginal increase beyond 
that is probably the signature of the Kaiser effect.
In what follows, the average systematic offset will be 
subtracted out of the PSC$z$ and IRAS 1.2 Jy dipoles to 
statistically  correct for rocket-like effects. 
The importance of the different error sources seems to 
vary with the distance. In the inner regions nonlinear effects
and intrinsic error in the reconstruction seem to dominate. 
Uncertainties in the filling procedures and Kaiser effect becomes 
dominant on larger scales. The former accounting for $\sim 40$ \%
error budget at a distance of 100 \hmpc, 
both in amplitude and direction.
The typical random error on the differential dipole,
$\alpha$, amounts to $ \sim 15$ \kms per shell of 10 \hmpc. 
This error will be accounted for in the likelihood analysis 
of \S 4 in the same way as the shot noise. 

\begin{figure}
\parbox{15 cm}{%
\epsfxsize=\hsize\epsffile{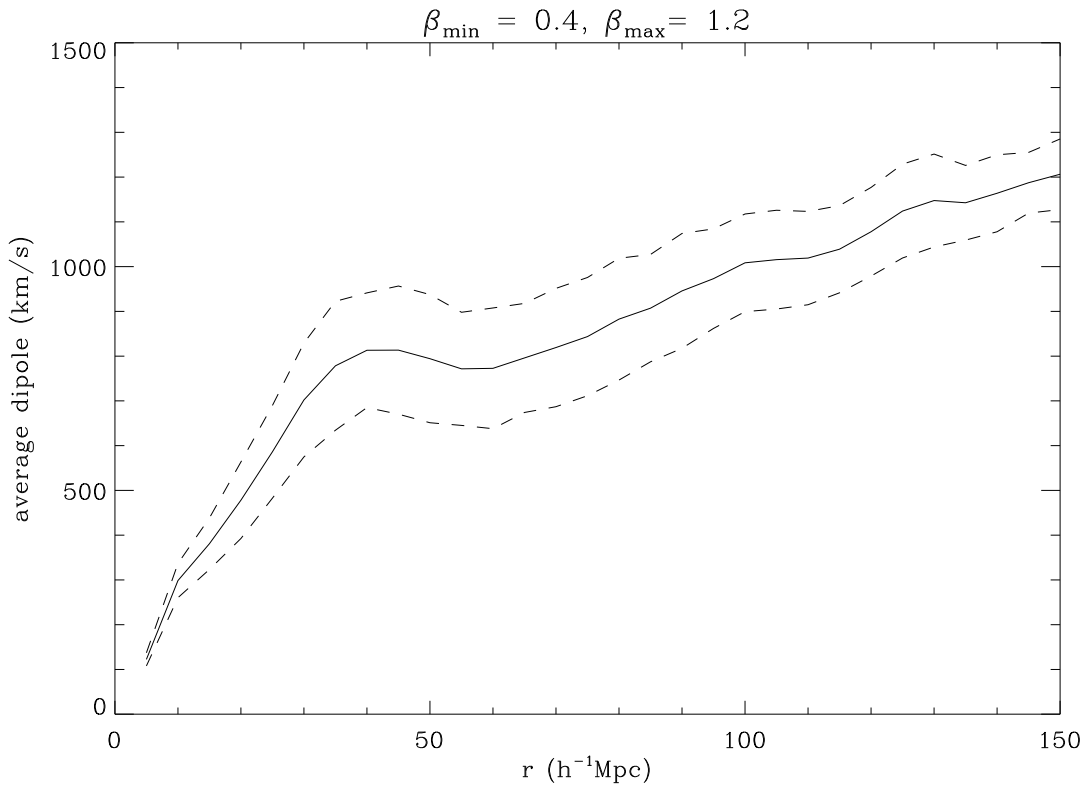}}
\caption{Real Space PSC$z$ cumulative dipole reconstructed 
with the M1 method  using different {\it a priori}
values of $\beta$ between 0.4 and 1.2. Solid line is the average of
all results, dashed lines represent the dipoles calculated with the
highest and lowest values for $\beta$.}
\end{figure}

The errors showed in figure 1 turned out to be Gaussian distributed
around their mean (i.e. the black dots) which indicates that our
error estimate, albeit based upon only 20 PSC$z$ mock catalogs,
are realistic.
As outlined in \S 3.2, the reconstructed dipole scales with $\beta$ 
linearly only to a first approximation. The actual result marginally
depends on the value of  $\beta$ assumed in the reconstruction itself.
To evaluate  the amplitude of this effect we have performed 9
PSC$z$ reconstructions varying $\beta$ in the range [0.4,1.2],
to include the range allowed by observations (see Dekel 1997).
Figure 2 shows the average cumulative dipole from the 9 iterations. The upper
and lower dashed lines display the dipoles for the 
$\beta=1.2$ and $\beta=0.4$, respectively. 
At the radius of $150 \hm$, the percentage
difference is 11 \%, notably smaller than the 32 \% caused by 
the cumulative effect of random errors and shot noise. 
Also, note that the shape of the dipole remains
constant at large radii which. This corroborates our previous 
conclusion that systematic errors induced by the Kaiser rocket effect 
are far less prominent than in the S92 analysis.

Uncertainties deriving from the  nonlinear $\beta$ scaling
will therefore be neglected in the rest of the work.

\subsection{The PSC$z$ Dipole} 
\label{sec:cumdip}

\begin{figure}
\parbox{15 cm}{%
\epsfxsize=\hsize\epsffile{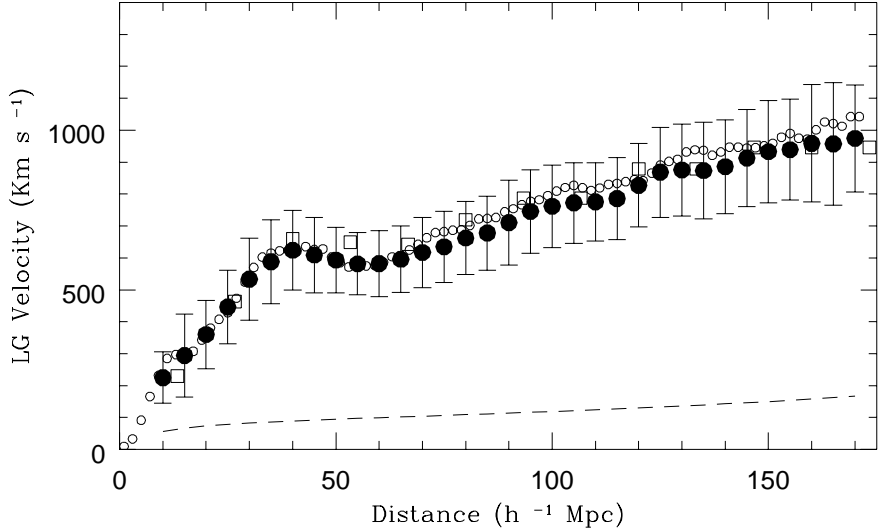}}
\caption{PSCz cumulative acceleration: the amplitude. Black circles
represent  
the amplitude, normalised to $\beta=1$,
of the cumulative PSCz dipole obtained when correcting
for redshift space distortions using method 1 (squares = method 2,
open circles = method 3). 
Error bars are the the same as in figure 1. 
The dashed line on the bottom represents the cumulative shot noise
error. No correction for systematic uncertainties has been applied.}
\end{figure}

We are now ready to compute the PSC$z$ dipole corrected for systematic 
biases.
Figure 3 shows the  magnitude of the 
cumulative PSC$z$ acceleration up to the limiting depth 
of $170 \hm$, computed after the reconstruction procedure but 
with no correction for systematic effects.
The amplitude is normalised to $\beta=1.0$.
The real space dipole computed using M1 is shown by filled circles
while the dipoles from M2 and M3 reconstructions are displayed 
with squares and open circles, respectively.
Error bars representing the same random errors as in figure 1 are 
displayed for M1 alone to avoid overcrowding. 
The cumulative dipoles of the three different 
reconstructions appear to be
remarkably similar, despite the very different nature of the
techniques used. 
The same considerations also apply for the dipoles' directions and 
speak for the reliability of our dipole estimate.

\begin{figure}
\parbox{15 cm}{%
\epsfxsize=\hsize\epsffile{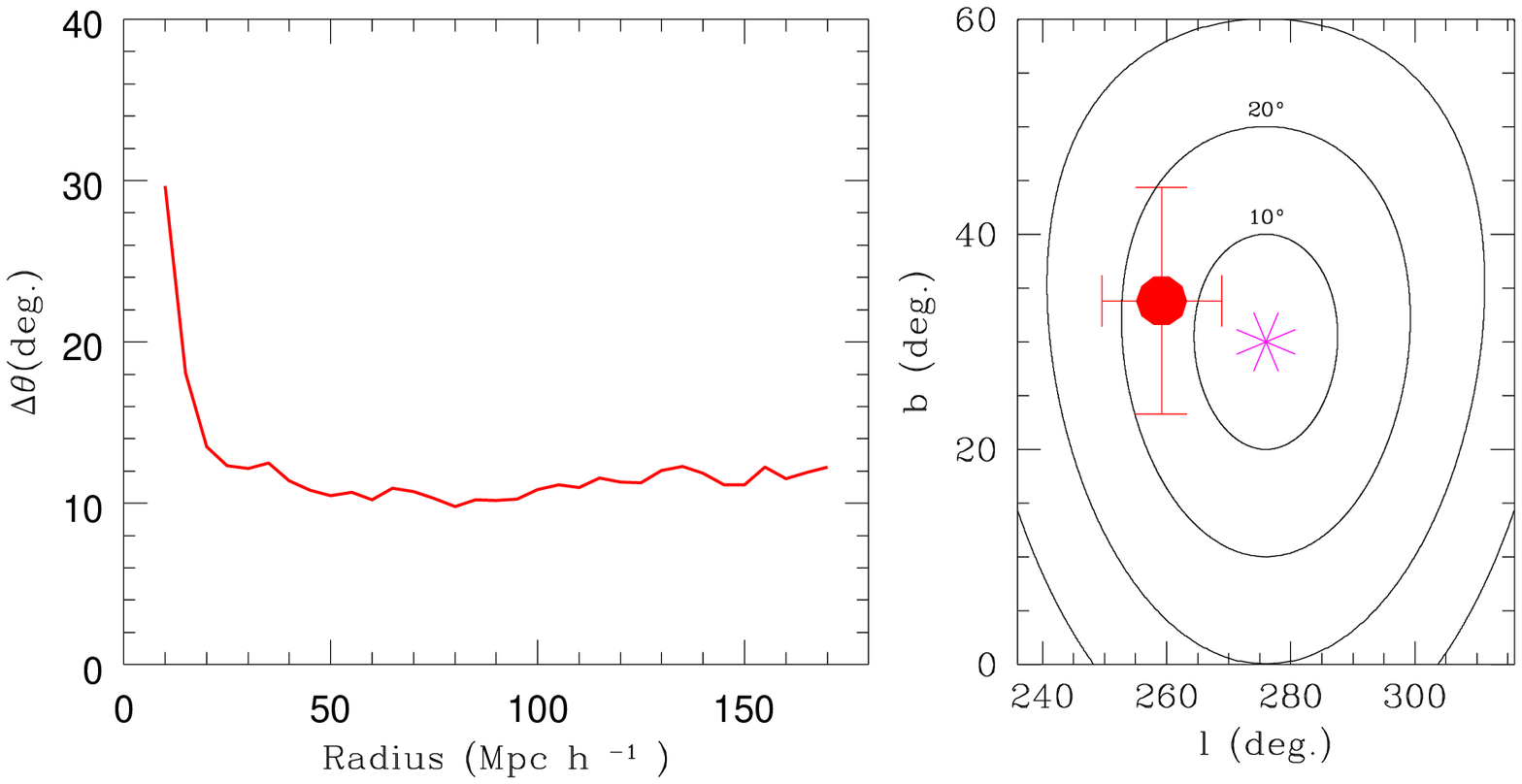}}
\caption{PSCz cumulative acceleration: the direction. 
The plot on the left shows the cumulative misalignment between 
PSCz and CMB dipole vectors. On the right the black dot represents the 
direction on the sky of the cumulative PSC dipole as measured at 150
\hmpc. Errorbars quantify 1--$\sigma$ random errors from the mocks. 
The starred symbol at the centre shows the CMB dipole directions. Contours are
drawn at constant misalignment angles.}
\end{figure}

Since the three methods produce equivalent results let us,
from now on, concentrate on the M1 dipole only.

\begin{table}
\centering
\tabcolsep 2pt
\begin{tabular}{lccccccc}  \\ \hline
& R (\hmpc) & $|\vv_r|$ (\kms) & $\sigma_{|\vv|}$ (\kms) &
 $l$ (deg.) & $b$ (deg.) & $\Delta\theta$ (deg.) &    \\ \hline 
& 10  &  224.8 & 80.4  & 252.1 & 68.7 &  29.7 & \\
& 20  &  360.1 & 107.0 & 252.0 & 53.5 &  13.5 & \\
& 30  &  533.0 & 128.8 & 271.6 & 42.6 &  12.2 & \\
& 40  &  624.0 & 124.2 & 269.2 & 36.9 &  11.9 & \\
& 50  &  593.2 & 102.7 & 266.9 & 36.2 &  10.5 & \\ 
& 60  &  582.1 & 103.2 & 263.9 & 30.4 &  10.2 & \\
& 70  &  617.0 & 110.2 & 263.3 & 29.7 &  10.8 & \\
& 80  &  662.4 & 114.7 & 265.1 & 33.5 &   9.8 & \\
& 90  &  710.5 & 132.4 & 265.2 & 36.5 &  10.2 & \\
& 100 &  761.5 & 129.6 & 262.4 & 33.7 &  10.9 & \\
& 110 &  775.7 & 122.8 & 261.3 & 33.9 &  10.9 & \\
& 120 &  827.4 & 130.7 & 260.5 & 34.0 &  11.3 & \\
& 130 &  875.5 & 144.0 & 257.6 & 34.5 &  12.0 & \\
& 140 &  885.7 & 147.4 & 259.4 & 33.0 &  11.8 & \\
& 150 &  932.8 & 159.6 & 259.2 & 33.8 &  11.1 & \\ 
\hline
\end{tabular}
\caption[]{Cumulative PSC$z$ Dipole. 
Column 1: Distance from the LG in ;
Column 2: Cumulative amplitude in ($\beta=1$);
Column 3: Random errors in ($\beta=1$);
Column 4: Cumulative direction along $l$ ;
Column 5: Cumulative direction along $b$ ;
Column 6: Cumulative misalignment w.r.t. the direction of the CMB
dipole.}
\label{tab:self1}

\end{table}
Table 1 shows the amplitude of the uncorrected cumulative
dipole and the shot noise contribution, both normalized to $\beta =1$,
along with the dipole direction along $l$ and $b$ and the 
cumulative misalignment angle of the reconstructed  
PSC$z$ dipole from the CMB dipole apex
The latter quantity is shown in the left panel of
figure 4. The misalignment is remarkably small already beyond $40 \hm$, 
i.e.  when large 
overdensities (Great Attractor and Perseus Pisces supercluster)
enter the sampled volume. 
This indicates that contributions to the LG accelerations 
are coherent over a very large range of scales (Basilakos and Plionis 
1998).
The right hand panel displays the 
direction of the cumulative gravity vector measured within $150 \hm$
while the error bars quantify the random uncertainties 
derived from the mocks.
The continuous curves are loci of constant misalignment in the $(l,b)$
plane. The asymptotic misalignment angle between LG and CMB dipole 
$\delta \theta=15^{\circ}$ is remarkably small.

\begin{figure}
\parbox{15 cm}{%
\epsfxsize=\hsize\epsffile{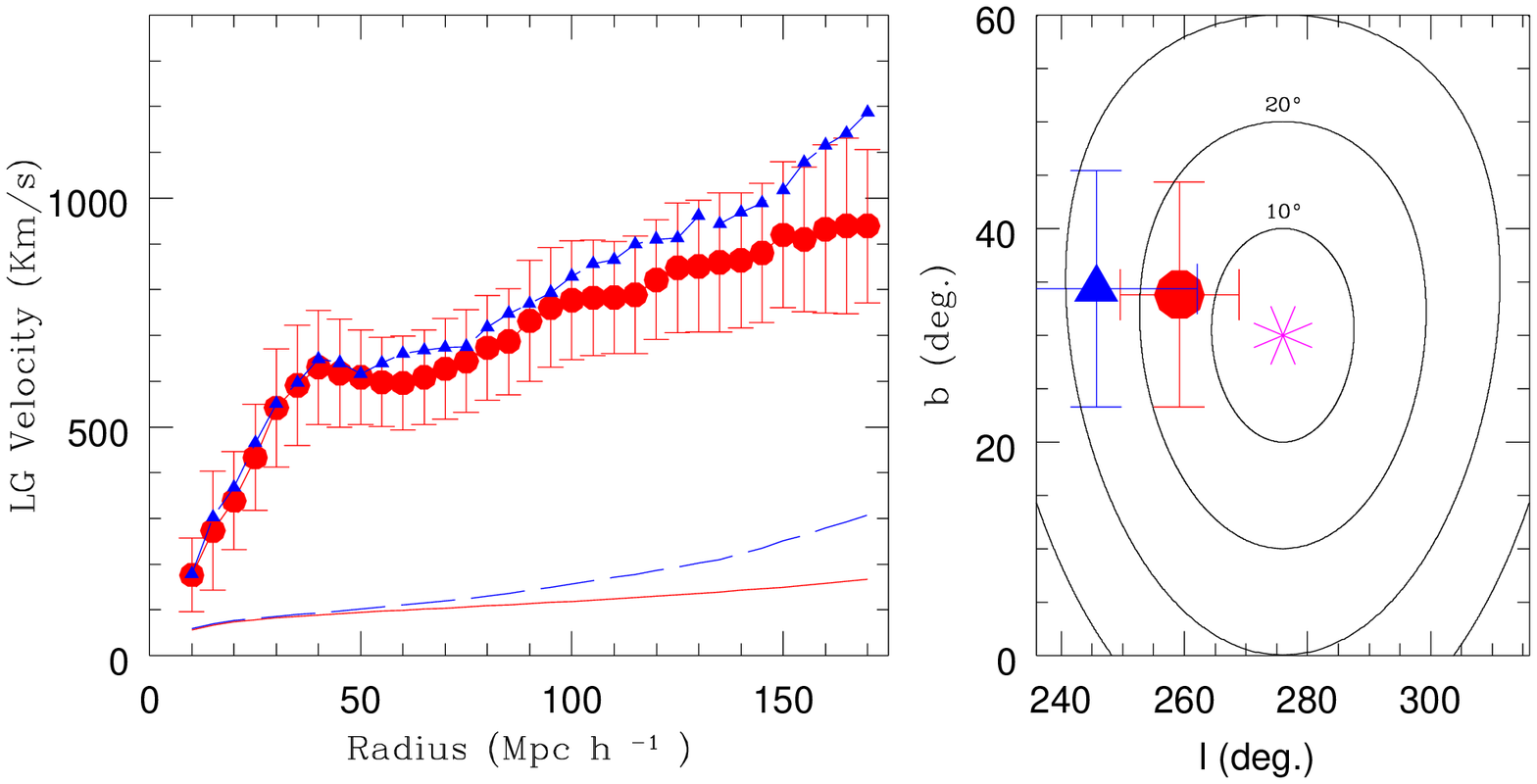}}
\caption{PSCz vs. 1.2 Jy dipoles. The amplitudes of the PSCz and
the 1.2 Jy dipoles, both
corrected for systematic errors, are shown in the left plot using
filled dots and  triangles, respectively. Error bars represent
1--$\sigma$ uncertainties. The lines on the bottom indicate the cumulative
shot noise error for the PSCz dipole (continuous line) and for the 1.2 
Jy one (dashed line).
On the right plot the same symbols indicate
the directions of the two dipoles at 150 \hmpc.} 
\end{figure}

Both PSC$z$ and 1.2 Jy samples have been extracted from the 
same parent PSC IRAS catalogue and have a similar sky coverage. 
The main difference between the 
two is the limiting flux which causes the radial selection
to be more severe for the IRAS 1.2 Jy galaxies. 
The dipoles from the two samples should therefore exhibit
similar features apart from different shot noise errors.
As shown in figure 5, we indeed find a remarkable agreement.
The left panel shows the cumulative  amplitudes of the two 
dipoles, both normalised to $\beta=1$.
Filled dots and small triangles refer to the PSC$z$ and 1.2 Jy samples, 
respectively.
Error bars represent random errors for the PSC$z$ dipole alone. 
Unlike in figure 4, dipoles in figure 5 have been statistically corrected 
for the systematic errors (as calculated in \S 3.4), which mainly
arise from the rocket effect and the  
uncertainties in the filling procedure.
The two curves on the bottom display the shot 
noise errors for the PSC$z$ (continuous line) and 1.2 Jy (dashed line).
Both corrected dipoles show the same features: a sharp rise followed
by a small decrease around 50 \hmpc 
as a result of the competing pull 
of the Great Attractor and Perseus Pisces regions.
There is a further slow rise in the PSC$z$ dipole of $\sim 35 \%$ 
in the range between [60-140] \hmpc.
Both the PSC$z$ and the 1.2 Jy show 
significant evidence for gravitational contributions beyond 
40 \hmpc. However, unlike the amplitude of the 1.2 Jy dipole, that of
the PSC$z$ seems to be constant beyond $\sim 140$ \hmpc.
This  discrepancy can be accounted for once 
the different shot noise contributions at large radii 
are taken into account.
The right hand side of figure 5 shows how large the dipole's misalignment 
becomes when 
using the 1.2 Jy catalogue. The 1.2 Jy gravitational acceleration within 
150 \hmpc 
points $\sim 25^{\circ}$ away from the CMB dipole, which reduces
to a mere $\sim 15^{\circ}$ when the PSC$z$ sample is considered (see
also Kolokotronis \etal (1996) who predicted the misalignment of the
PSC$z$ dipole to lie between 10 and 17 degrees). 
Error bars represent random errors -- no correction for
systematic effects is needed as they mainly 
affect the dipole's amplitude.

A further proof of the robustness of our result comes from the
comparison with preliminary results from the analysis of the PSC$z$ dipole 
by Rowan-Robinson \etal (1998), which used a different 
methodology and a much deeper PSC$z$ 
subsample to address the issue of its convergence. 
The two dipole estimates agree well within 1--$\sigma$ errors.
Also  note that the above results are fully consistent 
the recent  re--analysis of the 1.2 Jy and QDOT catalogues by 
Basilakos and Plionis (1998).

\subsection{A Numerical Estimate of $\beta$}
\label{sec:rob}

In this section, we will evaluate $\beta$ in a purely numerical way to
compare to the likelihood analysis. The  idea is similar to that of \S
3.4, i.e. we use the mock catalogues to evaluate the errors in the
reconstructed dipole. Here mock PSC$z$ catalogues are used to
statistically correct for the discrepancy between the measured
acceleration vector and the true peculiar velocity of an observer.
%E This has been added to account for two Strauss' criticism.
In this exercise we are assuming that the two cosmological models
assumed to generate the mock catalogues are realistic.
Indeed, the two cosmologies
explored produce  large scale motions compatible with the
observational constraints (Jenkins \etal 1998).
Moreover, for the cosmological models we have considered,
the contribution to the dipole from the 
wavelength longer than the size of the N--body computational box
is small compared to the others errors. Even more so in the 
inner regions, where a very significant contribution to the 
LG acceleration comes from. Therefore we do not account for
this error source in the present analysis. 
 
For each mock catalogue we compute, at different radii, the ratio between the
reconstructed dipole at the observer's position and its true
N--body velocity. The average of the ratios calculated from the
various mock catalogues provides us with a multiplicative
factor which relates the dipole reconstructed by our method
to the real LG velocity. We calculate this for all three Cartesian
components 
and find that the multiplicative factors are sufficiently similar to
warrant averaging. 

Note that, unlike in \S 3.4, the discrepancy between reconstructed 
dipole and real velocity is caused by 
the effects of non-linearity, shot noise, finite volume and residual
redshift space distortions
i.e. this now represents the
cumulative effect of {\it all} sources of error. 
Note also that the multiplicative factor and its dispersion do not depend on
$\beta$ in the first approximation when it has been averaged over all
catalogues. The reconstructed dipoles have
been scaled to what is in each case known to be the correct $\beta$
of the simulation
so that the ratio between the reconstructed and the real dipole velocity only
depends on the various errors and not on cosmology.

\begin{figure}
\parbox{15 cm}{%
\epsfxsize=\hsize\epsffile{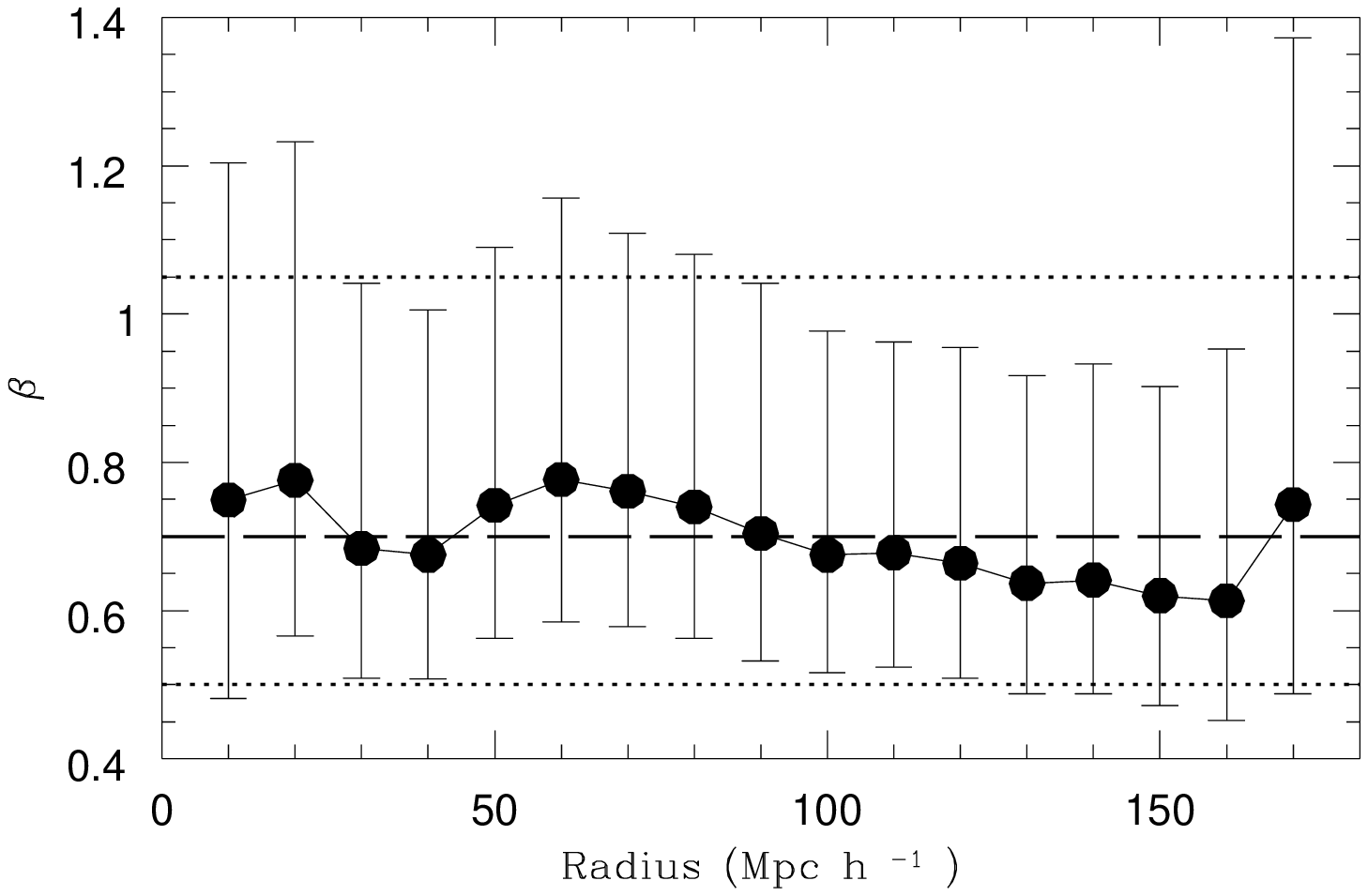}}
\caption{Robustness Test. Filled dots represent the $\beta$ value
obtained when comparing the LG velocity deduced from the CMB dipole 
to the one numerically deduced from the PSCz acceleration.
Errorbars represent 1--$\sigma$ uncertainties
estimated from the mock PSCz catalogs. 
The dotted and the dashed lines represent the 1--$\sigma$ level 
range for the value of beta from the likelihood analysis 
($\beta=0.7_{-0.2}^{+0.5}$).}
\end{figure}

We then use this multiplicative factor to recover 
the LG peculiar velocity 
from the reconstructed cumulative PSC$z$ dipole to be compared 
with CMB dipole, $v_c$, to evaluate $\beta$. 

Figure 6 shows the result of this
procedure. Surprisingly, $\beta$ is remarkably stable around 0.7 even at low
radii, where the volume includes little of the
mass distribution  that really causes the total  LG acceleration. 
However, the errorbars on those
scales are correspondingly larger. 
The errors, which represent the dispersion in the distribution 
of cumulative factors obtained from the mocks, 
are asymmetric, which
indicates that it will be easier to set a lower rather than an upper
limit on $\beta$.

\begin{figure}
\parbox{15 cm}{%
\epsfxsize=\hsize\epsffile{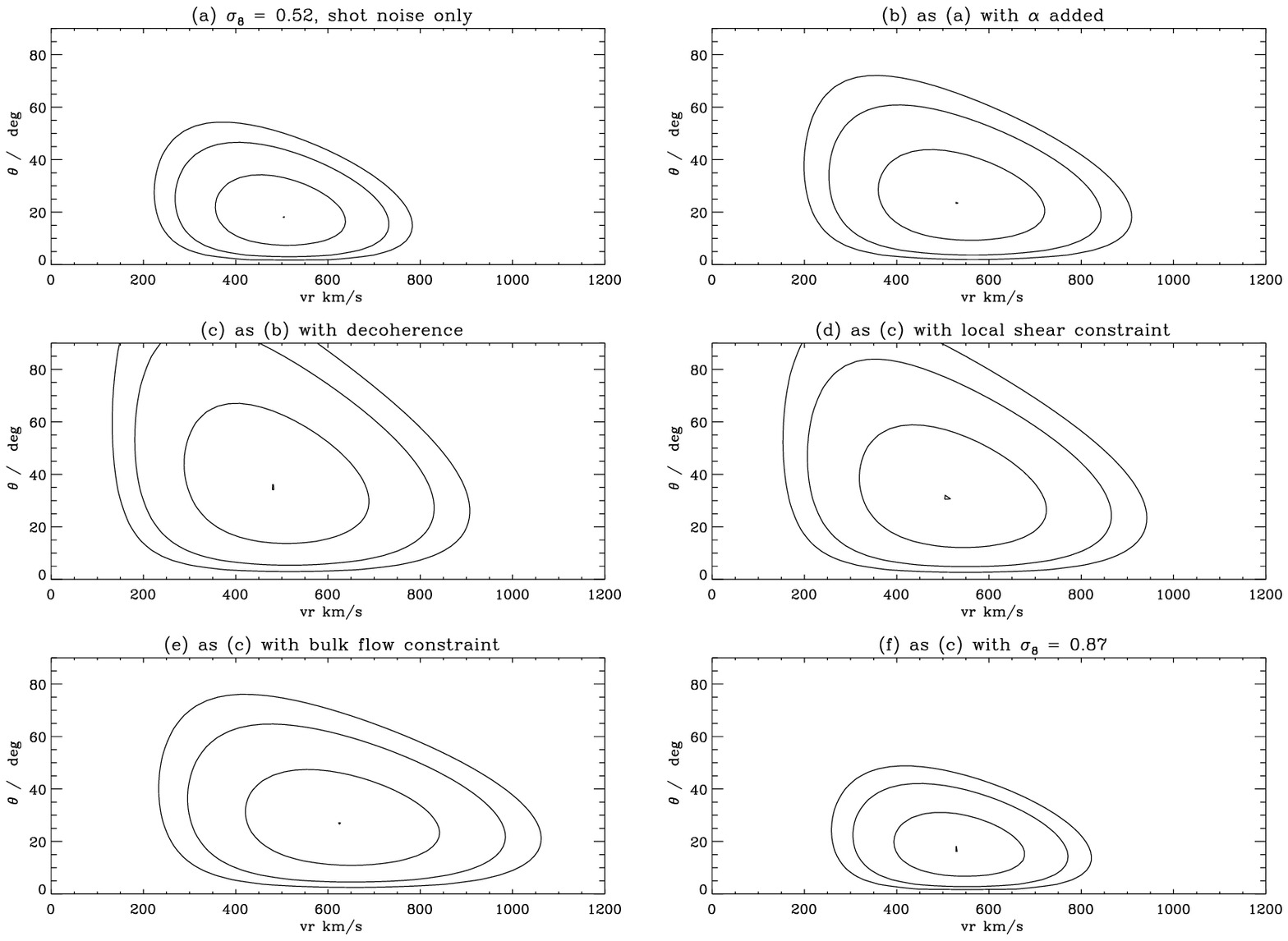}}
\caption{Probability of Amplitude and Misalignment. 
Likelihood contours 
for amplitude and misalignment
of the PSCz dipole in a specified world model. Fig 7a refers to the
case of  ($\Gamma = 0.5$,$\beta =1$,$\sigma_8=0.52$) universe
while contours refer to ${\cal L}(\bv_r \mid \bv_{CMB})$. Fig 7b show
the effect of introducing the reconstruction errors, $\alpha$, 
and Fig 7c also includes the
decoherence effects. Fig 7d and e show the effect of introducing either
the shear or the bulk flow constraints, respectively. 
Figure f show the
same case as figure 7d for a 
($\Gamma = 0.195$, $\beta=0.55$ and $\sigma_8 =0.87$) model.}
\end{figure}

The dashed and two dotted lines show the result of the likelihood
analysis performed in \S 4.3 (figure 7a and 7b).
The dashed line represents the most 
likely $\beta$-value, the dotted
lines the $\beta$ $1 - \sigma$ range.
It is encouraging that the numerical analysis is entirely consistent with
the likelihood analysis. 

\section{Likelihood Analysis}
\label{sec:LA}

\subsection {Velocity Statistics}
\label{sec:VS}

Our aim is to assess the likelihood of a given cosmological
model using all the information we have on velocities in the local
universe, i.e. the measured LG velocity, the CMB dipole, the bulk
velocity, and the LG velocity residual.
To this end, we need to find a mathematical expression for the
probability distribution functions (joint or constrained) of the
velocities available and the first step is to find expressions for these
velocities. 
In the present analysis, which is based on the 
work of Juszkievicz, Vittorio and Wise (1990) and S92,
all velocities are regarded as Gaussian random
vectors. 

In the limit of a finite survey volume, the peculiar velocity of the
Local Group becomes
\begin{equation}
\bv_i= \frac{H_{\circ} \beta}{4 \pi} \int d^{3} \br \prime
W_i(r)\frac{\br\prime}{\mid \br\prime\mid^{3}}
\delta_{g}(\br\prime) 
\label{lgvelwr}
\end{equation}
where the window function $W_i(r)$ specifies the finite survey
volume.
It is often more convenient to work in Fourier space in which
case eqn. (\ref{lgvelwr}) becomes
\begin{equation}
\bv_i= H_{\circ} \beta\int d^{3} \bk
\tilde{W}_i(k)
\frac{i \ \bk}{(2 \pi)^3 k^2}
\delta_g(\bk),
\label{lgvelwk}
\end{equation}
where $\bk$ is the wavenumber vector,  $\tilde{W}_i(k)=k\int dr 
W_i(r)j_1(kr)$,
is the window function in Fourier space and 
$j_1(kr)$ is the spherical Bessel function of order 1. 
This 
expression represents the 
linear velocity of the LG that is expected to be measured through a
given window function in a universe characterised by the density field
$\delta_g$. 
To make use of the information stored in 
the growth of the cumulative LG acceleration vector 
with radius of the sampled volume, we  will find it more 
convenient to measure the differential LG velocity, $\bv_{r,i}$
generated within a
series of $N$ non-overlapping top-hat window functions between radii $r_i$
and $r_{i+1}$,
\begin{equation}
\tilde{W_i}(k) = \frac{\sin(kr_i)}{kr_i} - \frac{\sin(kr_{i+1})}{kr_{i+1}}
\label{wi}
\end{equation}
with $i=2, ... N$.
In the innermost region ($i=1$) we smooth with a top hat filter
on scales smaller than 
$r_s = 5 \hm$ to eliminate strong nonlinear effects.
Therefore, the window function  for the innermost region is of the form
\be
\tilde{W_1}(k) = \frac{3 j_1(k r_s)}{kr_s} - j_o(k r_{1}).
\label{onewindow}
\ee
The total measured LG velocity $\bv_r$
can be obtained when $N=1$ and $r_{1}=R_{max}$ is the sample's depth. 

The other velocities modelled in this analysis are equally
characterised by different window functions.
The $N+1$th velocity vector considered 
is the true LG velocity measured in the CMB frame, $\bv_c$. 
Its window function 
extends over all space except for a small--scale cut--off to account for 
the
finite size of the Local Group. In $k$--space its form is
\begin{equation}
\tilde{W}_{CMB} = \frac{\sin(kr_{LG})}{(kr_{LG})},
\end{equation}
where $r_{LG} = 1 \hm$. 

The velocity field within the neighbourhood of the Local Group 
is remarkably quiet (\cf Peebles 1998). The observed LG velocity residuals
$\bv_s$, i.e.
the difference between the Local Group 
velocity and the average velocity within a sphere of $5 \hm$, is less 
than 
200 \kms. This can be used as an effective constraint in a likelihood 
analysis of the LG dipole field and we take the 
amplitude of the velocity residuals
as the $N+2$th random variable.
The window function for the 
LG velocity residuals 
within $r_a = 5 \hm$ is
\begin{equation}
\tilde{W}_{s}(k)  = 1 - 3 \frac{\sin(kr_a) - kr_a
\cos(kr_a)}{(kr_a)^3}.
\end{equation}

The final $N+3$th random vector we consider is the bulk velocity
$\bv_b$ measured within $r_b = 30 \hm$ from the POTENT analysis 
of the Mark III catalogue of galaxy peculiar velocities (Dekel 1997).
Since we consider POTENT--processed data the measured 
bulk velocity is just the average peculiar velocity of a spherical region
of $30 \hm$. This allows us to express the relative 
window function for the bulk flow as
\begin{equation}
\tilde{W}_b(k) = 3 \frac{\sin(kr_b) - kr_b
\cos(kr_b)}{(kr_b)^3}.
\end{equation}

Since we model all the $N+3$ velocities as Gaussian random vectors, any
probability distribution function will only depend on the covariance
matrix:
\begin{eqnarray}
M_{lm} &=& \frac{1}{3} < {\bf v}_l \cdot {\bf v}_m > \nonumber \\
       &=& \frac{H_{\circ}^{2} \beta^{2}}{6 \pi^2} 
\int dk P(k) \tilde{W}_l(k) \tilde{W}_m(k), 
\label{covariancematrix}
\end{eqnarray}
where $P(k)=\langle \delta(k)\delta(k) \rangle$ is the power spectrum of 
density fluctuations and $\tilde{W}_l(k)$, $\tilde{W}_m(k)$ are any of
the window functions that we have previously introduced.
In this paper, we restrict our analysis to the family of CDM
power spectra for which we use the expression of Davis \etal (1985),
 characterised by the shape factor $\Gamma=\Omega_m h$:
\begin{equation}
P(k) = \frac{B k \exp[-\frac{1}{2}(kg_s)^2]}{[1+1.7k/\Gamma
+ 9(k/\Gamma)^{3/2} + (k/\Gamma)^2]^2}
\end{equation}
where $g_s = 100 \kms$ and the only function of the exponential term
is to improve numerical stability.
$h$ is the Hubble constant in units of 
$100 \kms$ Mpc$^{-1}$ and $B$ is the normalisation constant. 
As anticipated in \S 3.3,
we normalise the power spectrum to 
$\sigma_8$ according to the Eke, Cole and Frenk (1996) prescription.
Thus, we can completely characterise the 
CDM background cosmology by specifying $\beta$, $\Gamma$ and 
$\Omega_{\Lambda}$.

\subsection {Probability Distributions}
\label{sec:lik}

The Gaussianity of the velocity field guarantees that 
the joint distribution function of all our N+3 model velocities
is also a multivariate Gaussian:
\begin{equation}
f({\bf v}_r(i),i = 1...N, {\bf v}_{c}, {\bf v}_s, \bv_b) = 
(2\pi)^{-3(N+3)/2} (det M)^{-3/2} \exp(-\frac{1}{2} {\bf v}_l \cdot
{\bf v}_m (M^{-1})_{lm}) d\bv_r d\bv_c d\bv_s d\bv_b
\label{fv}
\end{equation}
where $l,m$ take values from 1 .. N+3 and Einstein summation convention
is assumed.
This represents the relative probability of observing any given set of these
velocities in some assumed CDM universe specified by ($\beta$,
.$\Gamma$) and the normalisation $\sigma_8$. 
To treat any of $\bv_c$, $\bv_b$, or $\bv_s$ as a constraint, we have
to calculate the ratio of the joint probability to the probability of
the constraint. For example, the probability distribution function of
$f(v_r(i), i = 1 .. N \mid v_c)$ is given by
\be
f(v_r(i), i = 1 .. N \mid v_c) = (2 \pi)^{-3N/2} \left(
\frac{M_{cc}}{det M} \right) \exp \left[ -1/2 \left( \bv_l \cdot \bv_m
(M^{-1})_{lm} - \frac{v_c^2}{M_{cc}} \right) \right] d\bv_r(i)
\ee
Any other constraint is similarly easy to express. 
For the CMB dipole vector we use the recent determination by Linewaver
\etal (1996) quoted in the introduction.
For the bulk flow vector we use the Mark III-POTENT estimate 
at $R_b = 30 $ \hmpc (437 \kms $\pm$ 40 \kms, l=306$^{\circ}$,
b=14$^{\circ}$ Dekel 1997). 
We have chosen not to use the 
Mark III-POTENT at large radii, which would set a more stringent constraint, 
because of the problems in the calibration of the distance indicators recently
discussed by Willick and Strauss (1998) beyond
30 \hmpc. Within a scale of 30 \hmpc, however, Mark III velocities
have proved to be reliably calibrated 
and fully consistent with the 1.2 Jy gravity field 
(Willick \etal 1997 Davis, Nusser and Willick 1996).

For the LG velocity residuals,
we impose a constraint on their magnitude $v_s \le 200 \vel$,
which means that the 
appropriate probability distribution has to be integrated 
over a sphere of radius $v_a = 200 \vel$. 
The expressions for the various conditional 
distribution functions involving the first $N+2$ vectors,
$ f({\bf v}_r(i)  \mid \ {\rm constraints})$, have all been 
derived by S92, and can be easily extended to include 
the bulk velocity vector.

By construction, 
the various probability distributions already account for errors arising
from the use of finite window functions. However, as we have already pointed 
out, there are other error sources that we need to account for explicitly:

\begin{itemize}

\item As mentioned above, we adopt the shot noise definition
of S92 (their eq. (35))  
which accounts for sparse sampling as well as  
the unknown mass function of the IRAS galaxies. 
The latter is quantified by the mass variance which for PSC$z$ galaxies,
in the S92 formalism, is $K=1$. 
As displayed in figure 3, the PSC$z$ shot noise  increases
with radius. We calculate the shot noise contribution to the
differential dipole in each shell, square it and add it to the
diagonal elements of the covariance matrix, $M_{ii}$, with $i \leq N$. 

\item The PSC$z$ dipole is measured from the 
real space position of galaxies which we obtain after minimising the 
redshift space distortions iteratively. Intrinsic errors
in the reconstruction procedure generate an average random error 
of 15 \kms in the differential dipole 
(\cf \S 3.4) produced by a shell of matter  
10 \hmpc wide. This error, $\alpha$, is added 
in quadrature to the
first $i=1...N$ diagonal elements of the covariance matrix just like
the shot noise.

\item Nonlinear motions spoil the alignment between $\bv_r$ 
and $\bv_c$. To account for these effects, we follow S92 and define 
a decoherence function in $k$--space which expresses the misalignment
between the Fourier components of the LG velocity 
$\bv_{r,k}$ and the CMB dipole $\bv_{c,k}$:
\be
W_d(k) = \frac{\langle \bv_{r,k} \cdot 
\bv_{c,k}\rangle}{|\bv_{r,k}||\bv_{c,k}|}.
\ee
This additional window function multiplies the integrands 
that describe the covariance between measured accelerations and LG
velocity, i.e. the terms $M_{ij}$ with $i \leq N$, $j = N+1$ and their
conjugates. Clearly, since nonlinear effects are more severe on small 
scales, $W_d(k) \rightarrow 0$ at large $k$. 
To model $W_d(k)$ we use the S92 expression 
\begin{equation}
W_d(k) = \frac{1}{(1+(kr_d)^4)^{\frac{1}{2}}}.
\label{dec}
\end{equation}
We put the decoherence length $r_d = 4.5 \hm$
which S92 templated using N--body simulation of a standard CDM
cosmology. In adopting this expression we follow the S92 assumption 
that the decoherence function does not change appreciably with the 
background cosmology.

\end{itemize}

>From the probability density distributions introduced
above and the measured PSC$z$ dipole, it is
possible to evaluate the relative likelihood,  ${\cal L}$, of different 
world models $(\beta, \Gamma, \sigma_8)$.
We define the likelihood function as
\begin{equation}
{\cal L} = -2 \ln(f),
\label{lf}
\end{equation}
which is conveniently distributed like $\chi^2$ around its minimum. 
In the next two sections we will perform two different likelihood
analyses with the aim of constraining the $\beta$ and $\Gamma$
parameters that, along with $\sigma_8$, specify the underlying
cosmological model. 

\subsection{Probability of Amplitude and Misalignment}
\label{sec:am}

The aim of this first section is to afford us some insight into
what kind of dipole we are likely to measure given
a model of the universe and certain constraints.
We do this by measuring the probability that an
LG--like observer carrying out a  PSC$z$--like survey measures a 
velocity $\bv_r$ in a given $(\beta, \Gamma,\sigma_8)$ universe.
We characterize the LG velocity by its amplitude measured
within a specified window, $v_r$, and the misalignment to the CMB
dipole $\theta$. Since we want to 
plot the values of 
the likelihood  distribution only as a function of $v_r$ and $\theta$,
we compute  the constrained probability
rather than the joint one.
It is worth noting that $v_r$ scales with the assumed $\beta$
and that the differential element in the probability distribution becomes
$ d\bv_r = \beta^3 v_r^2 \sin(\theta) dv_r d\theta d\phi $.

Figure 7 shows 68 \%, 90 \%, and 95 \% likelihood 
contours from  six probability distributions. In all cases 
the acceleration is measured within the same spherical window function 
of radius $150 \hm$ which means that we will be using the window
function of  eqn.  (\ref{onewindow}) in which $R_{max}=150$
\hmpc. 
In figure 7a we plot
${\cal L}(\bv_r \mid \bv_{c})$ for a universe with 
$\Gamma = 0.5$, $\beta =1$ and $\sigma_8=0.52$,
only accounting for shot noise
errors (it can been shown that in the absence of  shot noise and 
with an infinitely deep window function the likelihood contours reduce to a 
point around $\theta = 0$ and $v_r = 620 \kms$).
The variance in the predicted direction and amplitude (i.e. the width
of the contour lines) is remarkable
which shows how misleading it can be to evaluate $\beta$ from a
direct comparison between $v_r$ and $v_c$.
It becomes slightly larger when  accounting for the errors in the
reconstruction procedure, $\alpha$, as shown in figure 7b.
The inclusion of the decoherence has an even more dramatic
effect on the likelihood contours. Figure 7c shows how
seriously our ignorance in modelling nonlinear motions can affect 
the determination of the true dipole. The effect of 
decoherence is to eliminate the large
$k$--modes which  greatly increases the variance in the measured dipole 
direction and brings the most likely misalignment up to $\sim 30^{\circ}$.
The LG velocity residuals constraint imposes an upper limit on the amplitude of
nonlinear contributions to peculiar velocities that originate on small
scales and thus should reduce the variance introduced by the
decoherence. As shown 
figure 7d, however, the effect on the likelihood contours is
quite minor.
Much better results are obtained when including the bulk flow
constraint, as is shown in 
figure 7e where we have replaced the velocity residuals
constraint with the bulk flow one. The variance in $\theta$ is greatly
reduced and the peak of the likelihood moves towards larger values of
$v_r$. This behaviour derives from imposing that $\vv_r$ and $\vv_c$
be well aligned which, together with the CMB dipole amplitude
constraint, forces a significant part of the LG acceleration to
originate within 30 $\hm$. 

Finally,  in figure 7f we have changed the background cosmology
$(\beta, \Gamma,\sigma_8)$. The plot shows the same case as figure 7c but
refers to a  $\sigma_8 =0.87$ normalization with 
$\Gamma = 0.195$ and $\beta=0.55$.
This model has smaller power on small scales,
which considerably reduces the spread of the likelihood contours.
The position of the central peak, however, 
does not change appreciably. This reflects 
the fact that 
CDM models normalised to the cluster abundance
produce large scale motions with similar properties
(Jenkins \etal 1997). In particular the most likely 
velocity of a LG--like observer remains remarkably constant.

\subsection{Probability of ($\beta,\Gamma$)}
\label{sec:bg}
\begin{figure}
\parbox{15 cm}{%
\epsfxsize=\hsize\epsffile{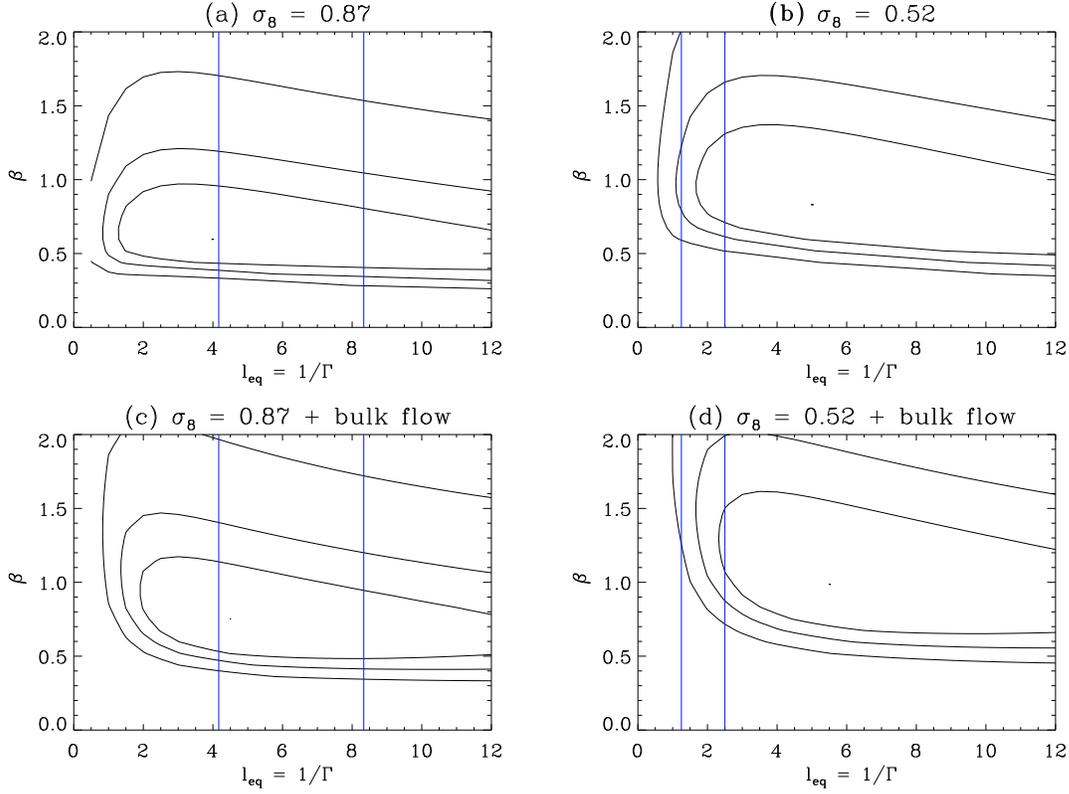}}
\caption{Probability of ($\beta,\Gamma$).
The plots 8a and 8b show the likelihood contours of ${\cal L}({\bf v}_r(i), i =
1,15,  {\bf v}_c \mid v_s < v_a)$, while the bulk flow constraint is
introduced in figures 8c and 8d. The case for a high  normalisation ($\sigma_8=0.87$)
is shown in figures a and c. A smaller value of $\sigma_8=0.52$ is used
to normalise the model plotted in figures b and d.
Vertical dashed lines delimitate the range in which $0.4<h<0.8$ as
suggested by observations. They have been drawn assuming that the
normalisation is meant to reproduce the cluster abundance.}
\end{figure}

In the second part of this analysis,
we consider a series  of different $(\beta, \Gamma)$ universes 
and compute, for each of them, the likelihood of obtaining the
same dipole as measured from the PSC$z$ subsample within a specified
window function and for a fixed $\sigma_8$ normalisation. 
The likelihood contours can then be displayed in a 
two--dimensional ($\beta$,$\Gamma$) plane.
Note that the Eke, Cole and Frenk
(1996) normalisation allows one to relate 
$\sigma_8$ to $\Omega_m$, which defines a 
$(\beta,\Gamma)\rightarrow(b,h)$ transformation. In the following, we
will, however, normally disregard this transformation and consider the
normalisation simply as a fixed parameter.
Also, allowing for a non-vanishing cosmological constant does not change 
the overall picture, as we have tested by comparing 
figure 8a with analogous contours produced  for an $\Omega_m=0.3$,
$\Omega_{\Lambda}=0.7$ scenario. We therefore limit our discussion to the case
of a $\Omega_{\Lambda}=0$ universe.

Figure 8 shows likelihood contours of $\beta,\l_{eq}=1/\Gamma$ for 
two different choices of $\sigma_8$ (we have chosen to use
the parameter $l_{eq}$ to allow easier comparison of our results to
those of S92).
Confidence levels in this case are 68, 90, 99 \%,
which translates into $\Delta {\cal L}$ of 2.3, 4.6, and 9.2 since
there are  2 degrees of freedom in the distribution. The function
plotted  is ${\cal L}({\bf v}_r(i), i = 1 ... N, {\bf v}_c \mid v_s <
v_a)$ in which we use
$N = 15$ non-overlapping top-hat windows of width $10$ \hmpc. This
makes use of the data we have on the growth rate of the dipole up to a
distance of 150 \hmpc, as plotted in figure 5.

Figure 8a,b show confidence contours of the function for both normalisations
($\sigma_8=0.87$ and $\sigma_8=0.52$ respectively).
Shot noise, intrinsic reconstruction errors, $\alpha$,
and decoherence effects
are all taken into account.
Note that with respect to the analysis of the IRAS 1.9Jy dipole 
by S92 (\cf their figures 13 and 14) our constraints on $\Gamma$ are 
somewhat weaker. 
This difference arises in part from the fact that, as we have
pointed out, the PSC$z$ dipole receives non-negligible contribution  
from mass 
inhomogeneities beyond 40 $\hm$, implying that models with large 
power (large $l_{eq}$) cannot easily be ruled out, whereas the data
used by S92 does not indicate any appreciable increase beyond 40
\hmpc. 
The two different contours in figures 8a and 8b
have basically the same shape, which
indicates that the analysis is not overly sensitive to the
normalisation of the power spectrum. Both exclude very small $\beta$
($< 0.4$) at the highest confidence level. This is not surprising
since a very small $\beta$ would not be able to produce the measured
velocities. The range of allowed $\beta$ is larger in the
low-normalisation case and slightly shifted towards higher values
because small fluctuations in the density contrast can only cause the
observed velocities when coupled with higher masses. In neither case
can we make any restrictive statement about $\Gamma$, except that the
power cannot be confined to the very smallest scales. The peak of the
likelihood distribution is shifted towards slightly higher powers in
figure 8b, again because if the fluctuations are small and the density is
generally larger, then power on large scales is more likely to be
responsible for the observed velocities. The peak of the likelihood
distribution in both cases is roughly consistent with the $\Gamma
\simeq 0.2$ measured by Tadros \& Efstathiou (1995) from the IRAS survey.

As mentioned above, the normalisation can be regarded as 
a constraint on $\Omega_m$ ($\Omega_m = 0.3$ for case 8a
and $\Omega_m = 1.0$
for 8b). In that case, the $\beta$- transforms to a b-axis
and the $\Gamma$- to a $h$-axis. In figures 8a,b the vertical lines
indicate the range of currently accepted $h$ ([0.4;0.8]), thereby
indicating an extra constraint on $\Gamma$ if $\Omega_m$ is considered
fixed. In both cases, this additional constraint is consistent with
the likelihood contours. In case 8b, however, it only allows fairly high
values of $\Gamma$, i.e. the 
high $\Omega_m$ (coupled with a low
normalisation to reproduce the observed cluster abundance) and a value
of $h$ in the given range force the power to originate on much smaller
scales than seems most likely from the measured velocities (or indeed
than was measured from the IRAS survey as mentioned above).

Figures 8c and 8d  show the effect of including the  bulk flow
constraint in figures 8a and 8b, respectively.
The bulk velocity is mainly determined by the shape and the amplitude of the
power spectrum on large scales. It is not surprising, therefore, that 
imposing the Mark III--POTENT bulk velocity constraint tends to favour
cosmological models with excess power on large scales. Since the bulk
flow is well aligned with the dipole, 
it also
excludes the combination of low $\beta$ and high $\Gamma$ in both
normalisations. In both cases, if the power is on small scales, the
observed bulk flow can only be generated with a very high
$\beta$. However, in figures  8c,d as well as in 8a,b, the likelihood analysis
cannot set stringent
limits on the shape parameter.
We only find that  $\Gamma < 0.4$ in case of low normalisation
and that $\Gamma < 0.5$ for the high-normalisation case.
Both limits refer to the 68 \% confidence limit level.

The likelihood analysis allows determination of 
$\beta$ with much better accuracy than $\Gamma$.
In the high-normalisation case we obtain $\beta=0.7_{-0.2}^{+0.5}$
with 68\% confidence. Note that this is the range allowed by the
likelihood contours in figure 8. We can tighten these constraints by
marginalising over the values of $\Gamma$ allowed by the
$h$-constraints. This yields $\beta = 0.7^{+0.35}_{-0.2}$ (see figure 9
for the marginalised probabilities corresponding to figure 8).
The resulting bias parameter for IRAS galaxies
is $b=1.43_{-0.2}^{+0.35}$, in accordance with 
the range of the values allowed by independent observations and
analyses. With $\sigma_8=0.52$ the likelihood contours shift 
upwards, increasing the most likely $\beta$ value
to 1.0$^{+0.6}_{-0.3}$ again at 68\% confidence level. 
\begin{figure}
\parbox{15 cm}{%
\epsfxsize=\hsize\epsffile{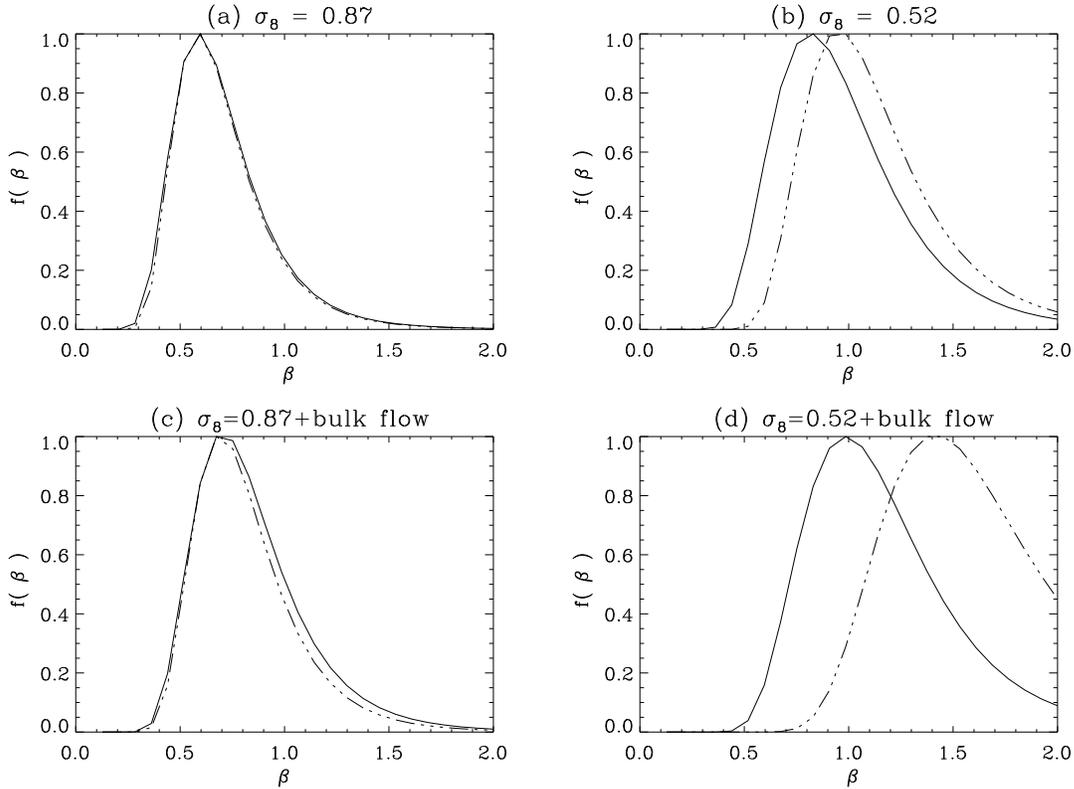}}
\caption{Marginalised likelihood distribution.
Marginal distribution of the likelihood of $\beta$ obtained 
from likelihood distribution. Plots a,b,c and d refer to their
analogous cases in figure 8. Continuous lines show the result 
of integrating over all the values of $\Gamma$. Dot dashed lines
are obtained when limiting the integration range between the vertical
dashed lines  displayed in figure 8.}
\end{figure}

\section{Conclusions}
\label{sec:conc}
 
We have measured the LG acceleration vector using the 
IRAS PSC$z$ galaxy distribution in real space.
Three independent methods have been used to correct for redshift space 
distortions and to keep systematic reconstruction errors under control.
As pointed out by S92, apart from the shot noise, the Kaiser
rocket effect represents the most serious uncertainty in the dipole 
measurement. In this work we have accounted for this and all the other
sources of systematic errors by using PSC$z$ mock
catalogues from N--body simulations.

The resulting PSC$z$ acceleration vector appears to receive a
non-negligible contribution from scales larger that 40 \hmpc, unlike 
in the IRAS 1.9 Jy (S92) and ORS (Strauss 1996)
analyses. We find that between 60 and 140 \hmpc the dipole amplitude 
increases by $\sim 35 \%$ and that only beyond this scale we find 
evidence for convergence. 
This result is in agreement with the recent
analysis by Rowan-Robinson \etal (1998) and also confirms
the reality of the large-scale contributions to the LG accelerations
advocated by Scaramella \etal (1991), Plionis \& Valdarnini (1991) and
Branchini \& Plionis (1996) on the basis of the distribution of the 
Abell--ACO clusters. The misalignment of the measured with the 
CMB dipole is remarkably
small ($\sim 15^{\circ}$) and stays almost constant beyond 40 \hmpc
which corroborates the impression that most of the LG acceleration 
is generated from density fluctuations along the Perseus Pisces--Great
Attractor--Shapley concentration baseline.
The IRAS 1.2 Jy dipole appears to be fully consistent with the PSC$z$
if the shot noise contribution is properly accounted for.
The increasing importance of shot noise and rocket effect beyond  the
depth of our mock catalogues leads us to limit 
our  PSC$z$ dipole analysis to within 170 \hmpc.

The measured PSC$z$ cumulative dipole has been fed into 
a likelihood analysis with the aim of constraining the $\beta$ parameter 
and the shape factor, $\Gamma$,
assuming a CDM framework and a given normalisation for the power 
spectrum.
The formalism accounts for shot noise errors, finite window size,
nonlinear effects and random uncertainties introduced when 
minimising redshift space distortions. 
To better determine $\beta$ and  $\Gamma$, observational constraints 
on the LG velocity residuals, bulk flow, 
and LG velocity relative to the CMB frame have been introduced in the
analysis.
We  account for the freedom in the spectrum normalisation by exploring two 
different cases: $\sigma_8=0.52$ and $\sigma_8=0.87$.
In both scenarios we find that the PSC$z$ dipole does not allow to set
stringent limits  
on the shape parameter $\Gamma$. 
However, if the normalisation is intended to reproduce the observed 
cluster abundance, then a relation between $\sigma_8$ and $\Omega_m$ results 
and further limits on $\Gamma$ can be set by constraining the Hubble
constant within the range allowed by observations.
As a result we find that in a critical universe (i.e. for
$\sigma_8=0.52$), $\Gamma$ is constrained to be between 0.4 and 1.0
which is well outside the range of 
recent $\Gamma$ estimates, like the one coming  from power
spectrum analyses of the IRAS survey (Tadros \& Efstathiou 1995)
for which $\Gamma \simeq 0.2$. 
It should however be noted that the
introduction of the constraints on $h$ implies that $\Gamma = \Omega_m
h$, and will not hold for more general expressions of the shape parameter.

The likelihood analysis produces more stringent constraints on
the $\beta$ value. 
For the high normalisation case (figure 8a) we find that
$\beta= 0.70^{+0.3}_{-0.2}$ at 1--$\sigma$ level. A result 
that does not change when the bulk flow constrain is introduced
(figure 8c). 
These results indicate that, if cluster normalisation is assumed,
then a low density universe
is very much consistent with observations and allows determination of 
the $\beta$ parameter with $\sim 35$\% accuracy .

When adopting a low normalisation,
the constraints over $\beta$ change appreciably with those on
$h$ and on the bulk flow.
In particular, they change drastically with $h$
when considering a low normalisation together with the bulk flow. The low
normalisation implies that $\beta$ has to be very high ($\beta \simeq
1.5$) to produce the
measured bulk flow (if $h$ is fixed within the given bounds). This is
much higher than $\beta$ determined from other
dynamical measurements. Together with the fact that the range in
$h$ also forces $\Gamma$ to be much higher than indicated by
independent measurements, this implies that the model is probably not
realistic. 

Our ability in determining $\beta$ from the galaxy dipole
is mainly hampered by 
our ignorance of the large scale contribution to the LG acceleration
by the fact that the likelihood analysis is based on 
one single observer. Better determinations of $\beta$ have 
been obtained by comparing MarkIII POTENT measured velocities to the 
ones modelled from the 1.2 Jy catalog
(Davis, Nusser and Willick, Willick et. al 1997, Willick and Strauss
1998, da Costa, Nusser et al 1998). 
A substantial improvement, however, can only be obtained 
by comparing the PSC$z$ gravity field with observed velocities at
many independent locations. An analysis of this kind
using the recent SFI data (Giovanelli \etal 1997a, 1997b)
is currently in progress.

\section*{Acknowledgments}
This work was supported by various PPARC grants and by the EC TMR network
for research in ``Galaxy formation and evolution.'' CSF acknowledges a
PPARC Senior Research Fellowship. LFAT was supported by the PRAXIX XXI
programme of JNICT (Portugal). IS and EB thank Michael Strauss 
for many useful comments and suggestions.

\end{document}